\documentclass[onecolumn,authoryear]{article}

\usepackage{amsmath,amssymb,amsfonts,amsthm,graphicx}
\usepackage{txfonts}
\usepackage{helvet}
\usepackage[utf8x]{inputenc} 
\usepackage{mathrsfs}
\usepackage{lmodern}
\usepackage[T1]{fontenc}
\usepackage{color}
\usepackage{authblk}
\usepackage{natbib}
\usepackage[hidelinks]{hyperref}

\newtheorem{definition}{Definition}
\theoremstyle{definition}

\newtheorem{remark}{Remark}
\theoremstyle{remark}

\AtBeginEnvironment{array}{\setlength{\arraycolsep}{1.2pt}}
\AtBeginEnvironment{eqnarray}{\setlength{\arraycolsep}{1.2pt}}

\DeclareMathAlphabet{\mathpzc}{OT1}{pzc}{m}{it}

\newcommand{\PAR}[2]{\frac{\partial{#1}}{\partial{#2}}}           
\newcommand{\MAP}[3]{{#1}:{#2}\to\mathbb{R}^{#3}}      
\newcommand{\col}{\mbox{col}}
\newcommand{\rank}{\mbox{rank}}

\def\rea{\mathbb{R}}

\begin{document}

\title{Passivity-Based Nonlinear Control}

\author{Pablo Borja}%


\affil{University of Plymouth, Plymouth, United Kingdom.}

\date{}

\pagestyle{empty}

\begin{center}
 \textbf{\Huge Passivity-Based Nonlinear Control}\\[2cm]
 {\Large Pablo Borja}\\[1cm]
 University of Plymouth, Plymouth, United Kingdom.
\end{center}

\vspace*{3cm}

\noindent\textbf{Disclaimer / Notice:} This is a preprint of an encyclopedia chapter accepted for 
publication in Encyclopedia of Systems and Control Engineering -- Volume 2, edited by Lorenzo 
Marconi, published by Elsevier. This version has not undergone final publisher proofreading or 
typesetting. The version of Record is available online via DOI:\\[0.15cm]
\url{https://doi.org/10.1016/B978-0-443-14081-5.00140-9}

\section*{Abstract}
The passivity-based control (PBC) framework focuses on understanding and modifying the energy 
storing and dissipation in the system to be controlled. To this end, PBC techniques often 
proceed in two steps: (i) ensuring that the closed-loop system's energy is minimum at the desired 
point, and then (ii) forcing the system to dissipate energy until reaching that point. These 
control methods have proven effective in controlling a wide range of systems, especially physical 
ones, even when they exhibit highly nonlinear behaviors.\\
This chapter discusses the main aspects of some PBC strategies for nonlinear systems.\\[0.15cm]
\textbf{Keywords:} Damping; Dissipation; Energy; Energy shaping; Interconnection; Lyapunov 
stability; Nonlinear systems;  Passivity; Physical systems; Port-Hamiltonian systems; 
Stabilization; Storage function.

\newpage

\maketitle

\textbf{Nomenclature}\\
\begin{equation*}
\begin{array}{ll}
\text{EL} &\text{Euler-Lagrange}\\
\text{IDA} &\text{Interconnection and damping assignment}\\
\text{PBC} &\text{Passivity-based control}\\
\text{PDE} &\text{Partial differential equation}\\
\text{pH} &\text{Port-Hamiltonian}\\
\text{PID} &\text{Proportional-integral-derivative}
\end{array}
\end{equation*}


\section{Introduction}
The term passivity-based control (PBC) was coined over thirty five years ago 
\citep{ortega1989adaptive}. Since then, the nonlinear control methodologies encompassed in this 
framework have proven suitable for controlling a wide range of systems---particularly physical 
ones---as shown in the abundant literature on this topic; see, for instance, the results reported 
in \cite{ORTbook,secchi,brogliato,GEObook,bai,VANJEL,VAN,PIDbook}. Some reasons for the success of 
PBC are the physical intuition of the strategies and the fact that, regardless of their complexity, 
the behavior of physical systems is ruled by their energy, which is, in turn, the main ingredient 
of PBC techniques. In particular, these approaches exploit the passivity property of dynamical 
systems, which, loosely speaking, means that the system cannot generate energy \citep{ORTcsm,VAN}. 
While this can be done in different ways depending on the approach, common ingredients of these 
nonlinear 
control strategies are energy shaping and damping injection. The former has its roots in the 
seminal work of \cite{takegaki} and involves assigning the desired minimum to the closed-loop 
system's energy. Furthermore, damping injection aims to ensure that the closed-loop system 
dissipates energy until the minimum is reached.

This chapter provides an overview of PBC for nonlinear systems in a state-space representation and 
discusses some recent and more specific advances regarding this nonlinear control framework.

\subsection*{Notation}
The symbol $I$ denotes the identity matrix. Similarly, the symbol $\mathbf{0}$ is reserved for 
vectors and 
matrices whose entries are zeros. The dimensions of $I$ and $\mathbf{0}$ follow from the 
context. Given $t\in\rea$, $x(t)\in\rea^{n}$, $\MAP{S}{\rea^{n}}{}$, and the mapping 
$\MAP{h}{\rea^n}{m}$, we adopt
the notation 
$$\nabla S(x) = \begin{bmatrix}
   \displaystyle\PAR{S(x)}{x_{1}} \\ \vdots \\ \displaystyle\PAR{S(x)}{x_{n}}
                \end{bmatrix}, \qquad \displaystyle\PAR{h(x)}{x}=\begin{bmatrix}
   \nabla h_{1}(x) & \dots & \nabla h_{m}(x)
\end{bmatrix},
\qquad \dot{x}=\dfrac{\mathrm{d}x(t)}{\mathrm{d}t}.$$
Given the (square) symmetric matrix $A\in\rea^{n\times n}$, the notation $A\succ 0$ indicates that 
$A$ is positive definite. Likewise, $A\succeq 0$, $A\prec 0$, and $A\preceq 0$ denote that $A$ is 
positive semi-definite, negative definite, and negative semi-definite, respectively. Given 
$x\in\rea^{n}$ and a positive (semi-)definite matrix $A\in\rea^{n\times n}$, $\lVert x \rVert$ and 
$\lVert x \rVert_{A}$ represent the Euclidean norm and a weighted Euclidean norm, respectively, 
i.e.,
$$\lVert x \rVert:=\displaystyle\sqrt{x^{\top}x}, \qquad  \lVert x 
\rVert_{A}:=\displaystyle\sqrt{x^{\top}Ax}. $$
Given the distinguished vector $x^{\star}\in\rea^{n}$ and the mappings $\MAP{S}{\rea^{n}}{}$ and 
$\MAP{f}{\rea^n}{n}$, we
define 
$$f^{\star}:=f(x^{\star}), \qquad \left( \nabla S \right)^{\star}:=\nabla S(x^{\star}).$$
Given the matrix $A\in\rea^{n\times n}$, $A_{i}$ denotes the $i^{th}$ column of $A$. The symbol 
$\col(\cdot)$ is a compact form to express a column vector, e.g., $\col(x_{1},x_{2}) = [ x_{1} \; 
x_{2}]^{\top}$.
\section{Passive Nonlinear Systems}\label{sec:passive}

Consider an input-affine nonlinear system decribed by the following expressions:
\begin{equation}\label{sys}
\begin{array}{rcl}
  \dot{x} &=& f(x(t))+g(x(t))u(t) \\
  y(t) &=& h(x(t))+j(x(t))u(t),
\end{array}
\end{equation}
where $t\in\rea_{\geq0}$ represents time;
$x:\rea_{\geq0}\to\mathcal{X}$, with $\mathcal{X}$ a differentiable $n$-dimensional 
manifold,\footnote{To simplify the presentation of the results, we assume that 
$\mathcal{X}\subseteq\rea^{n}$, unless something different is stated.} denotes the state vector;
$u:\rea_{\geq0}\to\rea^{m}$ denotes the input to the system, with $m\leq n$; 
$y:\rea_{\geq0}\to\rea^{m}$
is the system's output; $\MAP{f}{\mathcal{X}}{n}$ is sometimes referred to as the drift vector; 
$\MAP{g}{\mathcal{X}}{n\times m}$ is the input matrix, whose rank equals $m$ for every $x(t)$ in 
$\mathcal{X}$; $\MAP{h}{\mathcal{X}}{m}$; $\MAP{j}{\mathcal{X}}{m\times m}$ is called the 
feedthrough term. The inner product between the input and output, i.e., $u^{\top}(t)y(t)$ 
represents 
the external power supply rate. We assume that the solution $x(t)$ of \eqref{sys} is unique, 
$u(t)$ and $y(t)$ are bounded, and $\int_{0}^{t}u^{\top}(\tau)y(\tau) d\tau$ is well-defined.

In general, passivity is an input-output property, i.e., we determine if the map $u\to y$ is 
passive. However, for systems admitting a state-space representation of the form \eqref{sys}, we 
have the following definition of passive system.

\begin{definition}
The system described by \eqref{sys} is said to be passive if there exists 
$S:\mathcal{X}\to\rea_{\geq0}$, referred to as storage function, satisfying
\begin{equation}
 S(x(t))\leq S(x(0))+\displaystyle\int_{0}^{t}u^{\top}(\tau)y(\tau) d\tau. \label{disin}
\end{equation}
If \eqref{disin} is satisfied, but $S(x(t))$ is not bounded from below, i.e., 
$S:\mathcal{X}\to\rea$, \eqref{sys} is said to be cyclo-passive.
\end{definition}
Any functions $S(x(t))$ and signals $y(t)$ satisfying the inequality \eqref{disin} are referred to 
as storage functions and passive outputs, respectively \citep{WIL}. Notice that, if the storage
function is differentiable, the inequality \eqref{disin} can be expressed as
\begin{equation}
 \dot{S} \leq u^{\top}(t)y(t), \label{ddisin}
\end{equation}
which implies that the power extracted from the system cannot exceed the power injected into 
it. In general, dealing with \eqref{ddisin} is much easier than using \eqref{disin} 
because we are not required to compute the system’s trajectories when working with the former 
\citep{VAN}. Moreover, \eqref{ddisin} has a close connection with Lyapunov theory. To illustrate 
this, suppose that there exists a solution of \eqref{sys}, given by $x^{\star}\in\mathcal{X}$, such 
that $S(x)$ is positive definite at $x^{\star}$, i.e.,
\begin{equation}\label{Spos}
S^\star = 0, \qquad S(x(t))>0; \;\ \forall x(t)\in\mathcal{X}-\{x^{\star}\}.
\end{equation} 
Setting the input as 
\begin{equation}\label{Sdi}
 u(t) = -K_{\tt di}y(t),
\end{equation} 
with $K_{\tt di}\in\rea^{m\times m}$ positive definite, yields
\begin{equation}\label{Snon}
 \dot{S}\leq -\lVert y(t)\rVert^{2}_{K_{\tt di}}\leq 0.
\end{equation} 
From \eqref{Spos} and \eqref{Snon}, $x^{\star}$ is a stable equilibrium 
point for the closed-loop system with Lyapunov function $S(x(t))$. Furthermore, using analysis 
tools like Barbalat's lemma or LaSalle's invariance principle---see \cite{KHA} for further 
details---it can be concluded that $y(t)$ tends to 
zero as time tends to infinity. Consequently, if  
\begin{equation}\label{detec}
 \displaystyle\lim_{t\to \infty}y(t) = 0 \implies \displaystyle\lim_{t\to\infty}x(t) =x^{\star},
\end{equation} 
the equilibrium $x^{\star}$ is asymptotically stable. A more general version of the rationale 
exposed above is discussed in \cite{BYR91}.

In PBC jargon, the process of ensuring that \eqref{Spos} is satisfied is known as energy 
shaping, and the input choice \eqref{Sdi} is an example of damping injection. From a physical 
perspective, energy shaping guarantees that the desired equilibrium is the minimum of the 
closed-loop system's energy. Then, damping injection refers to the process of making the system 
dissipate energy until it reaches such an equilibrium.

\textbf{Caveat:} In the subsequent sections,  to simplify the notation, the argument $t$ is removed 
from all signals.

\subsection{Port-Hamiltonian Systems}
The port-Hamiltonian (pH) modeling framework has proven suitable for representing a broad range of 
systems \citep{GEObook,VANJEL}. From a PBC perspective, pH systems are convenient because the 
energy, interconnection pattern, and dissipative terms are explicitly provided. Furthermore, these 
systems are (cyclo-)passive, where the energy is a differentiable storage function. Because of 
this, 
many PBC results consider the closed-loop system, and sometimes also the open-loop one, to be pH 
systems. Below, we briefly recall some fundamental aspects of pH systems for the sake of 
completeness.

An input-state-output pH system is given by 
\begin{equation}\label{phsys}
\begin{array}{rcl}
 \dot{x} &=& \left[J(x)-R(x)\right]\nabla H(x)+g(x)u \\
 y &=& h(x) + j(x)u
\end{array}
\end{equation}
where $x$, $u$, and $g(x)$ are defined in the same way as for \eqref{sys}; the Hamiltonian 
function $\MAP{H}{\mathcal{X}}{}$ represents the total---virtual or physical---energy of the 
system; $\MAP{J}{\mathcal{X}}{n\times n}$ is the interconnection matrix, which is skew-symmetric; 
$\MAP{R}{\mathcal{X}}{n\times n}$ is the damping matrix, which is positive 
semi-definite---consequently, symmetric; $h(x)$ and $j(x)$, with 
the dimensions specified in \eqref{sys}, guarantee that $y$ is a passive output associated with 
$H(x)$.
To simplify the notation, we define the matrix
\begin{equation*}
 F(x):= J(x)-R(x).
\end{equation*}
Notice that \eqref{phsys} is a particular case of \eqref{sys}, where $f(x) = F(x)\nabla H(x)$. 
Furthermore, 
\begin{equation*}
 \left(\nabla H(x)\right)^{\top}f(x) = \left(\nabla H(x)\right)^{\top}F(x)\nabla H(x) = -\lVert 
\nabla H(x) \rVert^{2}_{R(x)}.
\end{equation*} 
Hence, as a consequence of Hill-Moylan's theorem \citep{hill76}, all the passive outputs 
associated with $H(x)$ are parameterized as follows:
\begin{equation}\label{paray}
\begin{array}{l}
 \begin{array}{rcl}
        h(x)&=& \left[g(x) + 2\phi^{\top}(x)w(x)\right]^{\top}\nabla 
H(x) \\ j(x)&=& w^{\top}(x)w(x)+N(x)
       \end{array}\\[0.35cm] \implies y = \left[g(x) + 
2\phi^{\top}(x)w(x)\right]^{\top}\nabla 
H(x) + [w^{\top}(x)w(x)+N(x)]u,
\end{array}
\end{equation} 
where $\MAP{N}{\mathcal{X}}{m\times m}$ is skew-symmetric; $\MAP{w}{\mathcal{X}}{\ell\times m}$, 
with $\ell\geq \rank\{R(x)\}$, is a free mapping; $\MAP{\phi}{\mathcal{X}}{\ell\times n}$ is a 
factor of the damping matrix, satisfying
\begin{equation*}
 R(x) = \phi^{\top}(x)\phi(x).
\end{equation*} 
Notice that $R(x)\succeq 0$ implies that $\phi(x)$ exists; nevertheless, it is not unique. We refer 
the reader to \cite{MENBOORT} for more details on the parameterization \eqref{paray}.

Most PBC results related to pH systems focus on the stabilization problem. Consequently, it is 
paramount to define the set of assignable equilibria for systems of the form \eqref{phsys}, which 
is given by
\begin{equation*}
 \mathcal{E}=\left\lbrace x\in\mathcal{X} \mid g^{\perp}(x)F(x)\nabla H(x) = \mathbf{0} 
\right\rbrace,
\end{equation*} 
where $\MAP{g^{\perp}}{\mathcal{X}}{(n-m)\times n}$ denotes the (full rank) left annihilator of 
$g(x)$, i.e., $g^{\perp}(x)g(x)=\mathbf{0}$ and $\rank\{g^{\perp}(x) \}=n-m$.
Notice that for any $x^{\star}\in\mathcal{E}$ exists $u^{\star}\in\rea^{m}$ such that
\begin{equation}\label{eqeq}
 F^{\star}(\nabla H)^{\star} + g^{\star}u^{\star} = \mathbf{0}.
\end{equation} 

\begin{remark}
 While pH models have proven suitable for representing many physical systems, models derived from 
the Euler-Lagrange (EL) formalism are more common in some disciplines, e.g., 
robotics. In most cases, it is possible to find an equivalence between an EL model and a pH 
one---see the discussion in \cite[Appendix B]{ORTbook}, \cite{jeltsema2009multidomain}, and 
\cite[Chapter 4]{VAN} for further details. Hence, the 
results presented in this chapter directly apply to most systems in the EL representation.  
\end{remark}
\section{Integrable Passive Outputs}\label{sec:int}

Passive outputs play a crucial role in PBC approaches. In particular, if the system to be 
controlled 
is (cyclo-)passive, integrable passive outputs can be exploited for energy-shaping purposes. 
Because 
the term ``integrable'' is an abuse of language in this context, we define the 
concept of integrable output for pH systems below.

\begin{definition}
 Given the pH system \eqref{phsys}, the output $y$ is said to be integrable if there exists 
$\MAP{\gamma}{\mathcal{X}}{m}$ such that
\begin{equation}\label{intcond}
 \dot{\gamma} = \left( \displaystyle\PAR{\gamma(x)}{x} \right)^{\top}\dot{x}=\left( 
\displaystyle\PAR{\gamma(x)}{x} \right)^{\top}\left[F(x)\nabla H(x)+g(x)u \right] = y.
\end{equation} 
\end{definition}

Independently of whether the passive output is integrable, damping can be injected with a 
controller of the form
\begin{equation}\label{gendi}
 u_{\tt di} = -\psi(y),
\end{equation} 
where $\MAP{\psi}{\rea^{m}}{m}$ satisfies
\begin{equation*}\label{psi}
 y^{\top}\psi(y)>0, \qquad \psi(y) =\mathbf{0} \iff y = \mathbf{0}.
\end{equation*} 
Notice that \eqref{Sdi} is a particular case of \eqref{gendi}, where $\psi(y) = K_{\tt di}y$.

\begin{figure}[h]
 \centering
 \includegraphics[width=0.9\textwidth]{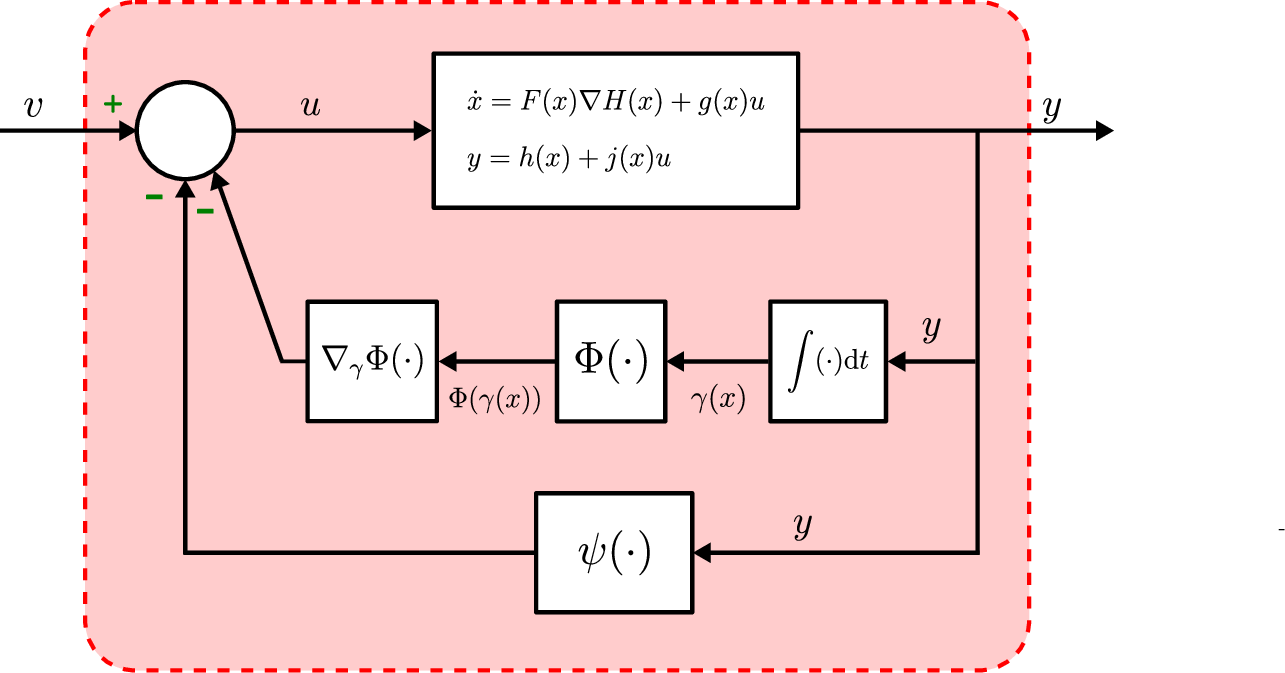}
 \caption{Schematic of \eqref{phsys} in closed-loop with \eqref{uinty}. The shaded area represents 
the resulting (cyclo-)passive system.}
 \label{fig:first}
\end{figure}

Given $\gamma(x)$ and $\psi(y)$, for any signal $v\in\rea^{m}$ and a
differentiable function $\MAP{\Phi}{\rea^{m}}{}$, the pH system \eqref{phsys} in closed-loop with
\begin{equation}\label{uinty}
 u = -\nabla_{\gamma}\Phi(\gamma(x))-\psi(y)+v
\end{equation}
yields another (cyclo-)passive with supply rate $v^{\top}y$ and storage function
\begin{equation}\label{Hdint}
 H_{\Phi}(x) = H(x)+\Phi(\gamma(x)).
\end{equation} 
To illustrate this, note that
\begin{equation*}
  \dot{H}_{\tt d} = \dot{H}+\dot{\Phi}  \leq y^{\top}u + 
\dot{\gamma}^{\top}\nabla_{\gamma}\Phi(\gamma(x))   = -y^{\top}\psi(y)+y^{\top}v  \leq
y^{\top}v.
\end{equation*}
Fig. \ref{fig:first} depicts the block diagram corresponding to the closed-loop 
\eqref{phsys}-\eqref{uinty}. In this approach, the function $\Phi(\gamma(x))$ shapes the energy of 
the new system, while the function $\psi(y)$ injects damping into it.

Suppose that the control objective is to stabilize a system of the form \eqref{phsys} at the 
desired equilibrium $x^{\star}\in\mathcal{E}$. Then, following the rationale exposed in Section 
\ref{sec:passive}, the controller \eqref{uinty}, with $v = \mathbf{0}$, achieves the control 
objective if $\Phi(\gamma(x))$ can be designed such that $H_{\Phi}(x)$ is---at least 
locally---positive definite at $x^{\star}$ and $y$ satisfies \eqref{detec}.

The role of each component of the control law \eqref{uinty} is clearly defined and has a 
physical interpretation, making this control strategy intuitive. Notably, some common control 
architectures for specific systems---for instance, the soft robots studied in 
\cite{BORDASAN}---can be understood as particular cases of this approach. Moreover, as shown in 
\cite{chan2023dead} and \cite{borja2023stabilization}, the functions $\Phi(\gamma(x))$ and 
$\psi(x)$ can be tailored to address implementation issues such as dead zones and saturation in 
the actuators. However, this approach faces two main challenges:
\begin{itemize}
 \item [(i)] Find integrable passive outputs.
 \item [(ii)] Given an integrable output $y$, ensure that $\Phi(\gamma(x))$ renders $H_{\Phi}(x)$
positive definite at $x^{\star}$.
\end{itemize}

While ensuring that every pH system has an integrable output is not possible, the
integrability of some passive outputs is easily verifiable. In particular, for pH systems with 
$F(x)$ invertible, the selection 
\begin{equation*}
w(x) = \phi(x)F^{-1}(x)g(x); \qquad N(x) = -g^{\top}(x)F^{-\top}(x)J(x)F^{-1}(x)g(x),   
\end{equation*} 
yields the passive output
\begin{equation}\label{yps}
 y = -g^{\top}(x)F^{-\top}(x)\dot{x}.
\end{equation} 
Recalling that there exists $\MAP{\gamma_{i}}{\mathcal{X}}{}$ such that
\begin{equation*}
 \nabla \gamma_{i}(x) = -F^{-1}(x)g_{i}(x)
\end{equation*} 
if and only if\footnote{The proof of this can be found in \cite{borjaUKACC}.}
\begin{equation}\label{poincare}
 \nabla (F^{-1}(x)g_{i}(x)) = \left[\nabla (F^{-1}(x)g_{i}(x)) \right]^{\top},
\end{equation} 
the output \eqref{yps} is integrable if and only if \eqref{poincare} holds for all the columns of 
$g(x)$. Consequently, a particular case in which \eqref{yps} is integrable occurs when $F(x)$ and 
$g(x)$ are constant matrices.
\begin{remark}
Suppose it is not possible to find integrable outputs associated with the storage function
$H(x)$. In that case, potential solutions are to change the input and output of the system via the
use of multipliers or to find a different storage function for the system.  These approaches are 
studied in \cite[Chapters 5 and 6]{PIDbook}. The mentioned reference also explains how to define 
an output similar to 
\eqref{yps} when $F(x)$ is not invertible. 
\end{remark}
Concerning challenge (ii), once an integrable output has been found, the existence of 
$\Phi(\gamma(x))$ rendering $H_{\Phi}(x)$ positive definite at $x^{\star}$ needs to be studied 
case by case. Nonetheless, similar to the integrability condition, the existence of 
$\Phi(\gamma(x))$ can be easily verified for some systems. To illustrate this, suppose that 
exists $\MAP{\bar\Phi}{\rea^{m}}{}$ such that
\begin{eqnarray}
 \left( \nabla H \right)^{\star} + \left( \nabla_{x} \bar{\Phi} \right)^{\star} &=&\mathbf{0} 
\label{gradint} \\
 \left( \nabla^{2} H \right)^{\star} + \left( \nabla^{2}_{x} \bar{\Phi} \right)^{\star} &\succ&0. 
\label{hessint}
\end{eqnarray} 
Hence, choosing
\begin{equation*}
 \Phi(\gamma(x)) = \bar\Phi(\gamma(x))-H^{\star}-\bar\Phi^{\star},
\end{equation*} 
makes the storage function $H_{\Phi}(x)$, defined in \eqref{Hdint}, locally positive definite at
$x^{\star}$. For pH systems satisfying \eqref{poincare}, the conditions \eqref{gradint} and 
\eqref{hessint} can be simplified. In particular, these systems have an integrable passive output 
given by \eqref{yps}. Moreover, because $F(x)$ is invertible and $x^{\star}\in\mathcal{E}$, 
\eqref{eqeq} can be expressed as
\begin{equation*}
 \left( \nabla H \right)^{\star} + (F^{\star})^{-1}g^{\star}u^{\star} = \mathbf{0}.
\end{equation*} 
Note that 
\begin{equation*}
 \nabla_{x}\bar\Phi(\gamma(x)) = -F^{-1}(x)g(x)\nabla_{\gamma}\bar\Phi(\gamma(x)).
\end{equation*} 
Thus, choosing $\bar\Phi(\gamma(x))$ such that
\begin{equation}\label{nphigs}
 \left( \nabla_{\gamma}\bar\Phi \right)^{\star} = -u^{\star}
\end{equation} 
we ensure that \eqref{gradint} is satisfied. We stress that $\bar\Phi(\gamma(x))$ satisfying 
\eqref{nphigs} always exists, where $u^{\star}$ can be computed as follows:
\begin{equation}\label{us}
 u^{\star} = -[(g^{\star})^{\top}g^{\star}]^{-1}(g^{\star})^{\top}F^{\star}\left( \nabla H
\right)^{\star}.
\end{equation} 
Now, suppose that $F(x)$ and $g(x)$ are constant matrices. Hence, \eqref{hessint} reduces to
\begin{equation}\label{hessint2}
 \left( \nabla^{2} H \right)^{\star} + F^{-1}g\left( \nabla^{2}_{\gamma} \bar{\Phi} 
\right)^{\star}g^{\top}F^{-\top} \succ0.
\end{equation} 
Because $\left( \nabla^{2}_{\gamma} \bar{\Phi} \right)^{\star}\in\rea^{m\times m}$ is a constant 
matrix, the existence of $\bar\Phi(\gamma(x))$ satisfying \eqref{hessint2} is guaranteed if there 
exists a symmetric matrix $K_{\tt es}\in\rea^{m\times m}$ such that
\begin{equation}\label{hessint3}
 F\left( \nabla^{2} H \right)^{\star}F^{\top} + gK_{\tt es}g^{\top}\succ 0.
\end{equation} 

From the analysis above, we can conclude that for pH systems of the form \eqref{phsys} with $F(x)$ 
and $g(x)$ full rank and constant, there exists $\Phi(\gamma(x))$ such that \eqref{gradint} and 
\eqref{hessint} hold if we can find a symmetric matrix $K_{\tt es}$ such that \eqref{hessint3} is 
satisfied. Indeed, under these circumstances, the function
\begin{equation*}
 \Phi(\gamma(x)) = \lVert \gamma(x)-\gamma^{\star} \rVert^{2}_{K_{\tt es}} - 
\gamma^{\top}(x)u^{\star} 
\end{equation*} 
ensures that $H_{\Phi}(x)$ defined in \eqref{Hdint} is locally positive definite at 
$x^{\star}\in\mathcal{E}$. Consequently, the controller \eqref{uinty}, with $v=\mathbf{0}$, ensures 
that $x^{\star}$ is a stable equilibrium for the closed-loop system. Furthermore, if the output 
\eqref{yps} satisfies \eqref{detec}, the closed-loop system has an asymptotically stable 
equilibrium 
at $x^{\star}$.
\section{PID Passivity-Based Control}\label{sec:pid}
This section focuses on PID-PBC results for pH systems. The chapter Introduction to Passivy-Based 
Control (see Section \ref{sec:cross}) provides an overview of PID-PBC for general passive systems; 
we refer the reader to that document for further details.

The main idea of PID-PBC is to construct a PID controller around the passive output such that the 
closed-loop system has a stable equilibrium at the desired point $x^{\star}\in\mathcal{E}$ 
\citep{PIDbook}. To this 
end, we can consider 
\begin{equation}\label{pidext}
 \begin{array}{rcl}
  \dot x_{\tt c} & = & y \\
u_{\tt pid} & = & -K_{\tt P} y - K_{\tt I} x_{\tt c} - K_{\tt D} \dot y,
 \end{array}
\end{equation} 
where the PID gains $K_{\tt P},K_{\tt 
I},K_{\tt D} \in \rea^{m \times m}$ satisfy
\begin{equation*}
 K_{\tt P}\succ 0, \qquad K_{\tt I}\succ0, \qquad K_{\tt D}\succeq 0.
\end{equation*}
Hence, a pH system of the form \eqref{phsys} in closed-loop with 
\begin{equation}\label{upid}
 u = u_{\tt pid}+v,
\end{equation} 
with $v\in\rea^{m}$, yields another (cyclo-)passive system with storage function
\begin{equation*}
 H_{\tt pid}(x_{\tt c},x) = H(x)+\frac{1}{2}\lVert y \rVert^{2}_{K_{\tt D}} + 
\frac{1}{2}\lVert x_{\tt c}\rVert^{2}_{K_{\tt I}}.
\end{equation*}
To show this, we compute the time derivative of $H_{\tt pid}(x_{\tt c},x,y)$, obtaining
\begin{equation*}
 \dot{H}_{\tt pid} \leq y^\top\left( u + K_{\tt D}\dot{y}+ K_{\tt I}x_{\tt c} \right)  = -\lVert y 
\rVert^{2}_{K_{\tt
P}}+y^{\top}v\leq y^{\top}v.
\end{equation*}
While \eqref{pidext} has the classical structure of a PID controller, and the closed-loop system is 
(cyclo-)passive, based on the analysis in Section \ref{sec:passive}, we require some additional 
conditions to guarantee the stability of the desired equilibrium $x^{\star}$. For instance, the 
closed-loop storage function satisfies the following relation
\begin{equation*}
 H_{\tt pid}(x_{\tt c},x) = 0 \iff \left\lbrace \begin{array}{rcl}
                                           H(x)+ \frac{1}{2}\lVert y \rVert^{2}_{K_{\tt D}}&=& 0 \\ 
x_{\tt c} &=& 
\mathbf{0}.
                                          \end{array}\right.
\end{equation*}
Accordingly, if $y$ vanishes at $x^{\star}$, $H_{\tt pid}(x_{\tt c},x)$ can only be positive 
definite at $(x_{\tt c}^{\star},x^{\star})$ if $x^\star$ is a critical point of $H(x)$. 
Nevertheless, the latter implies that $x^{\star}$ must be an equilibrium for the open-loop system, 
severely limiting the class of plants that could be stabilized using this approach. On the other 
hand, if $y$ is different from zero at $x^{\star}$, then closed-loop systems keeps dissipating 
energy at the desired equilibrium, implying that the trajectories of the closed-loop system cannot 
converge to $x^{\star}$; see \cite{ORTcsm,MENetal}. One way to circumvent these problems is to 
impose the integrability condition \eqref{intcond} on the output $y$. Hence, \eqref{pidext} can be 
reformulated as
\begin{equation}\label{pidy}
 u_{\tt pid} = -K_{\tt P} y - K_{\tt I} \left[\gamma(x)+\kappa\right] - K_{\tt D} \dot y,
\end{equation} 
where the constant vector $\kappa\in\rea^{m}$ is defined as follows:
\begin{equation}\label{kappa}
 \kappa: = -\gamma^{\star}-K_{\tt I}^{-1}u^{\star},
\end{equation} 
with $u^{\star}$ defined as in \eqref{us}.
Therefore, the storage function associated with the closed-loop system takes the form
\begin{equation*}
 H_{\tt pid}(x) = H(x)+\frac{1}{2}\lVert y \rVert^{2}_{K_{\tt D}} + 
\frac{1}{2}\lVert \gamma(x)+\kappa\rVert^{2}_{K_{\tt I}}.
\end{equation*} 
Notaby, if $y$ vanishes at $x^{\star}$, $\kappa$ given in \eqref{kappa} ensures that 
the control law \eqref{pidy} satisfies
\begin{equation*}
 u_{\tt pid}^{\star} = -K_{\tt I}(\gamma^{\star}+\kappa)  = u^{\star}.
\end{equation*} 
Consequently, $x^{\star}$ is an equilibrium for the closed-loop system \eqref{phsys}-\eqref{pidy}. 

The following remark establishes a relationship between the controllers exposed in Section 
\ref{sec:int} and PID-PBC.
\begin{remark}\label{int2PI}
 Setting $K_{\tt D}=\mathbf{0}$, the control law \eqref{upid}-\eqref{pidy} is a particular case of 
\eqref{uinty}, with
\begin{equation*}
 \Phi(\gamma(x)) = \frac{1}{2}\lVert \gamma(x)+\kappa\rVert^{2}_{K_{\tt I}}; \qquad \psi(y) = 
K_{\tt P}y.
\end{equation*} 
\end{remark}

Some well-posedness conditions must be met to guarantee that the controller 
\eqref{upid}-\eqref{pidy} can be implemented. To illustrate this, we consider $v=\mathbf{0}$; 
hence, $u = u_{\tt pid}$, with $u_{\tt pid}$ defined as in \eqref{pidy}. If $y$ 
has relative degree zero---i.e., $j(x)\neq\mathbf{0}$---and $K_{\tt D}=\mathbf{0}$, the 
controller reduces to
\begin{equation*}
 u =  -K_{\tt P}[h(x)+j(x)u]- K_{\tt I}
\left[\gamma(x)+\kappa\right],
\end{equation*} 
which can rewritten as
\begin{equation*}
 \left[I + K_{\tt P}j(x) \right]u = -K_{\tt P}h(x)-K_{\tt I}\left[\gamma(x)+\kappa\right].
\end{equation*} 
Consequently, to avoid singularities, the rank condition 
\begin{equation*}
 \rank\left\lbrace I+K_{\tt P}j(x)\right\rbrace = 
m
\end{equation*} 
must be satisfied such that the controller can be implemented as follows:
\begin{equation*}
 u = -\left[I + K_{\tt P}j(x) \right]^{-1}\left[K_{\tt I}\left[\gamma(x)+\kappa\right]+ K_{\tt 
P}h(x)\right].
\end{equation*} 
Similarly, to construct a suitable derivative term in PID-PBC---i.e., $K_{\tt D}\neq 
\mathbf{0}$---two conditions must be met: the controller must not require 
differentiation of $u$ nor exhibit singularities. The first condition implies that $y$ must have 
relative degree one---i.e., $y=h(x)$---while the second condition is satisfied if
 \begin{equation}\label{Kdcond1}
  \rank\left\lbrace I+K_{\tt D}\left( \displaystyle\PAR{h(x)}{x} \right)^\top g(x)\right\rbrace = 
m.
 \end{equation} 
Hence, if $y$ has relative degree one and \eqref{Kdcond1} is satisfied, the controller can be 
implemented as follows:
\begin{equation*}
 u = -\left[I+K_{\tt D}\left( \displaystyle\PAR{h(x)}{x} \right)^\top g(x)\right]^{-1}\left[K_{\tt 
P}h(x)+K_{\tt I}[\gamma(x)+\kappa]+K_{\tt D}\left( \displaystyle\PAR{h(x)}{x} 
\right)^\top F(x)\nabla H(x) \right].
\end{equation*} 
\begin{remark}
 In PID-PBC, the integral and derivative terms shape the energy of the system, while 
the proportional term injects damping into it. Moreover, the integral term, particularly $\kappa$, 
is central role in assigning the desired equilibrium.
\end{remark}

A major drawback of the analysis above is that adding a derivative term restricts the PID-PBC 
approach to passive outputs with relative degree one. Nevertheless, passive outputs with relative 
degree zero---for instance, \eqref{yps}---may be necessary to guarantee the output's integrability 
and, consequently, the construction of the integral term. To solve this problem, we can decompose 
the controller into two steps: (i) the formulation of an integral term based on a passive output 
with relative degree zero, and (ii) the addition of the proportional and derivative terms 
formulated using a passive output with relative degree one. To illustrate this idea, suppose that 
\eqref{yps} is an integrable output of \eqref{phsys}. Under these circumstances,
\begin{equation*}
 \displaystyle\PAR{\gamma(x)}{x} = -F^{-1}(x)g(x).
\end{equation*}
Hence, the control law
\begin{equation*}
 u = -K_{\tt I}[\gamma(x)+\kappa]+u_{\tt pd}
\end{equation*} 
yields
\begin{equation}\label{sysI}
 \dot{x} =  F(x)\nabla H(x) - g(x)K_{\tt I}[\gamma(x)+\kappa] +g(x)u_{\tt
pd} = F(x)\nabla H_{\tt I}(x) +g(x)u_{\tt pd}, 
\end{equation} 
with
\begin{equation}\label{HI}
 H_{\tt I}(x): = H(x) + \frac{1}{2}\lVert \gamma(x)+\kappa\rVert^{2}_{K_{\tt I}}.
\end{equation} 
Notice that
\begin{equation*}
 \dot{H}_{\tt I} = \left( \nabla H_{\tt I}(x) \right)^{\top}F(x)\nabla H_{\tt I}(x) + \left( \nabla 
H_{\tt I}(x) \right)^{\top}g(x)u_{\tt pd}  \leq \left( \nabla 
H_{\tt I}(x) \right)^{\top}g(x)u_{\tt pd}. 
\end{equation*} 
Accordingly, the system \eqref{sysI} has a passive output given by
\begin{equation*}
 y_{\tt I}= g^{\top}(x)\nabla H_{\tt I}(x).
\end{equation*} 
Moreover, $y_{\tt I}$ has relative degree one. Thus, if \eqref{Kdcond1} is satisfied for $h(x) = 
y_{\tt I}$, the PID architecture is completed by proposing
\begin{equation*}
 u_{\tt pd} = -K_{\tt P}y_{\tt I} - K_{\tt D}\dot{y}_{\tt I}.
\end{equation*} 
Therefore, from \eqref{sysI}, the closed-loop takes the form
\begin{equation}\label{clpid1}
 \begin{array}{rcl}
  \dot{x}  &=& F(x)\nabla H_{\tt I}(x)-g(x)\left(K_{\tt 
P}y_{\tt I} + K_{\tt D}\dot{y}_{\tt I}\right)\\[0.1cm] &=& \left[F(x)-g(x)K_{\tt 
P}g^{\top}(x)\right]\nabla 
H_{\tt I}(x)-g(x)K_{\tt D}\left( \displaystyle\PAR{y_{\tt I}}{x} \right)^{\top}\dot{x}.
 \end{array}
\end{equation} 
Hence, defining
\begin{equation}\label{K}
 \mathsf{K}(x): =  I +  g(x)K_{\tt D}\left( \displaystyle\PAR{y_{\tt I}}{x} 
\right)^{\top},
\end{equation}
we can rewrite \eqref{clpid1} as follows:
\begin{equation}
 \mathsf{K}(x)\dot{x} =  \left[F(x)-g(x)K_{\tt P}g^{\top}(x)\right]\nabla 
H_{\tt I}(x).
\end{equation} 
Thus, considering
\begin{equation}\label{Hdpidzero}
 H_{\tt pid}(x) = H_{\tt I}(x) + \frac{1}{2}\lVert y_{\tt I} \rVert^{2}_{K_{\tt D}},
\end{equation} 
we obtain
\begin{equation}\label{nHpid}
 \begin{array}{l}
  \nabla H_{\tt pid}(x) = \nabla H_{\tt I}(x) + \displaystyle\PAR{y_{\tt I}}{x}K_{\tt D}y_{\tt I} = 
\nabla H_{\tt I}(x) + \displaystyle\PAR{y_{\tt I}}{x}K_{\tt D}g^{\top}(x)\nabla H_{\tt I}(x) 
\\[0.25cm]\implies   \nabla H_{\tt pid}(x) = \mathsf{K}^{\top}(x)\nabla H_{\tt I}(x).
 \end{array}
\end{equation} 
Consequently, if the matrix $\mathsf{K}(x)$ has full rank, the closed-loop system admits the pH 
representation
\begin{equation*}
 \dot{x} = \mathsf{F}(x)\nabla H_{\tt pid}(x)
\end{equation*} 
with $H_{\tt pid}(x)$ given in \eqref{Hdpidzero} and 
\begin{equation}\label{Fpid}
 \mathsf{F}(x): = \mathsf{K}^{-1}(x)\left[F(x)-g(x)K_{\tt P}g^{\top}(x)\right]\mathsf{K}^{-\top}(x).
\end{equation} 

Notice that compared to the control approach exposed in Section \ref{sec:int}, the controllers 
obtained via PID-PBC have an extra component shaping the energy of the closed-loop system resulting 
from the derivative term. The following subsection illustrates how to exploit this term for 
stabilizing mechanical systems.

\subsection{PID-PBC for Mechanical Systems}
The total energy of a mechanical system is given by the Hamiltonian
\begin{equation*}
 H(q,p) = T(q,p)+V(q),
\end{equation*}
where $q\in\rea^{n}$ and $p\in\rea^{n}$ represent the generalized coordinates and corresponding 
momenta, respectively; $\MAP{V}{\rea^{n}}{}$ denotes the potential 
energy; $\MAP{M}{\rea^{n}}{n\times n}$ is the mass inertia matrix, which is positive definite; 
$\MAP{T}{\rea^{n}\times \rea^{n}}{}$ represents the kinetic energy, which is defined as follows:
 \begin{equation*}
  T(q,p):=\frac{1}{2}p^{\top}M^{-1}(q)p.
 \end{equation*} 
Given the Hamiltonian $H(q,p)$, an unconstrained mechanical system can be expressed 
as\footnote{In this case, $x=\col(q,p)\in\rea^{2n}$.}
\begin{equation}\label{mecsys}
 \begin{bmatrix}
  \dot{q} \\ \dot{p}
 \end{bmatrix} = \begin{bmatrix}
                  \mathbf{0} & I \\ -I & -D(q,p) 
                 \end{bmatrix}
\begin{bmatrix}
 \nabla_{q} H(q,p) \\ \nabla_{p} H(q,p)
\end{bmatrix}
+\begin{bmatrix}
  \mathbf{0} \\ G(q)
 \end{bmatrix}u;
\end{equation} 
where the input matrix $\MAP{G}{\rea^{n}}{n\times m}$ satisfies $\rank\{ G(q) \} = m$, and the 
$\MAP{D}{\rea^{n}\times \rea^{n}}{n\times n}$ is positive semi-definite. Some simple computations 
show that the set of assignable equilibria for \eqref{mecsys} is
\begin{equation}\label{eqmec}
 \mathcal{E} = \left\lbrace (q,p)\in\rea^{n}\times \rea^{n} \mid G^{\perp}(q)\nabla 
V(q)=\mathbf{0}, \; p=\mathbf{0} \right\rbrace.
\end{equation} 

Suppose that
\begin{equation*}
 G = \begin{bmatrix}
         \mathbf{0} & I
        \end{bmatrix}^{\top}; \qquad G^{\perp} = \begin{bmatrix}
              I & \mathbf{0}
             \end{bmatrix}.
\end{equation*}
Hence, the coordinates can be explicitly divided into actuated and unactuated, i.e.,
\begin{equation*}
 \begin{array}{lll}
  q_{\tt u} = G^{\perp}q, & \quad p_{\tt u} = G^{\perp}q, & \qquad q_{\tt u},p_{\tt u} 
\in\rea^{n-m}, \\
  q_{\tt a} = G^{\top}q, & \quad p_{\tt a} = G^{\top}q, & \qquad q_{\tt a},p_{\tt a} \in\rea^{m},
 \end{array}
\end{equation*} 
where the subscripts ${\tt a}$ and $\tt u$ stand for actuated and unactuated. Notice that under 
these circumstances
\begin{equation*}
 \dot{H} = \left( \nabla_{p}H(q,p) \right)^{\top}G u = p^{\top}M^{-1}(q)G u = \dot{q}^{\top}G u = 
\dot{q}_{\tt a}^{\top}u. 
\end{equation*} 
Therefore, $y = \dot{q}_{\tt a}$ is a passive output. Moreover, this output is integrable, with 
$\gamma = q_{\tt a}$. Consequently, \eqref{mecsys} in closed-loop with \eqref{uinty} yields a new 
(cyclo-)passive system with storage function
\begin{equation*}
 H_{\Phi}(q,p) = H(q,p) + \Phi(q_{\tt a})  = T(q,p) + V_{\tt 
d}(q), 
\end{equation*} 
with $V_{\tt d}(q):=V(q)+\Phi(q_{\tt a})$, implying that the term 
$-\nabla_{\gamma}\Phi(\gamma(x))$ in \eqref{uinty} modifies the potential energy of the system. 
Similarly, $\psi(y)$ injects damping into the actuated coordinates. Hence, from the discussion in 
Remark \ref{int2PI} we conclude that the integral and proportional terms in \eqref{upid} shape the 
potential energy and modify the system's damping, respectively.

\begin{figure}[h]
 \centering
 \includegraphics[width=0.7\textwidth]{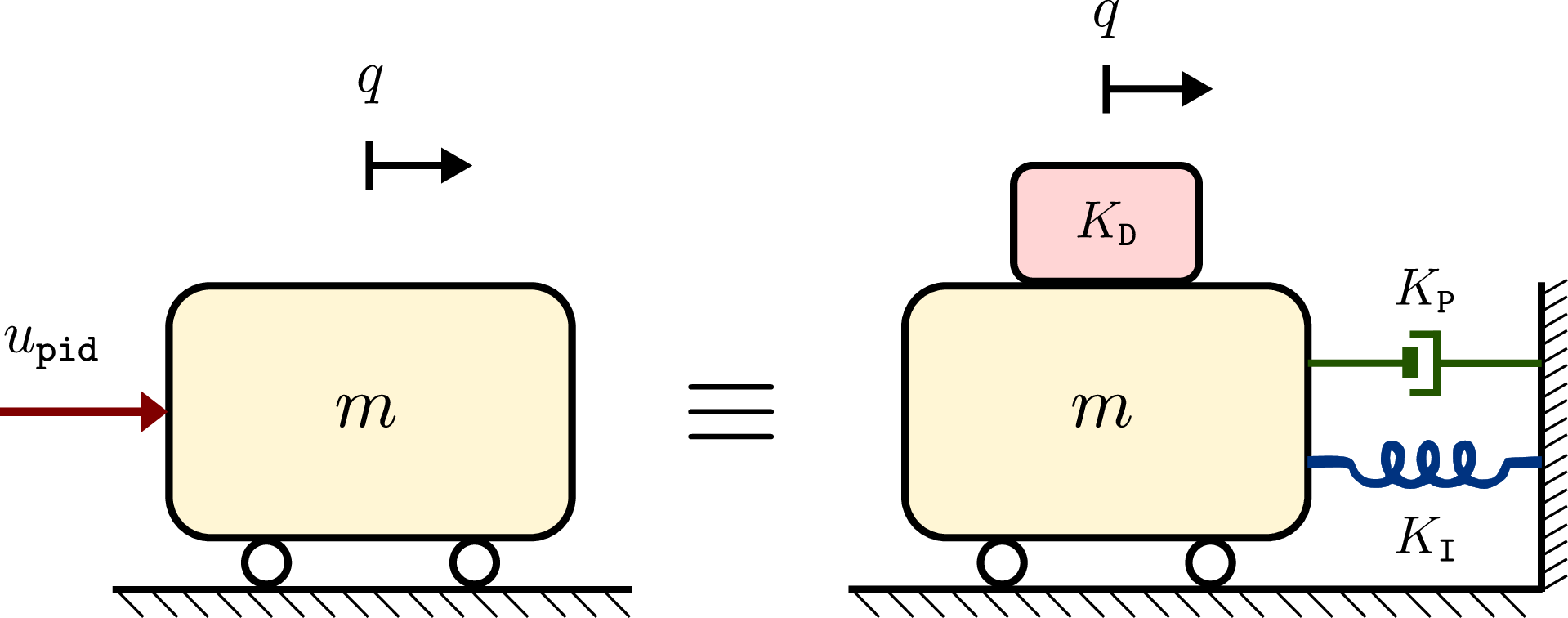}
 \caption{Physical interpretation of the effect of \eqref{pidy} on a mass. The integral 
term is interpreted as a spring that modifies the potential energy; the proportional term is 
understood as a linear damper; the derivative term is regarded as additional mass, changing the 
system's kinetic energy. In this schematic, $\kappa = 0$.}
 \label{fig:pid}
\end{figure}

Given \eqref{eqmec}, an assignable desired equilibrium for \eqref{mecsys} has always the 
structure $x^{\star} = \col(q^{\star},\mathbf{0})\in\mathcal{E}$. Accordingly, $T(q,p)$ is always 
positive 
at the desired equilibrium, implying that $H_{\Phi}(q,p)$ is positive definite at $x^{\star}$ if 
$V_{\tt d}(q)$ is positive definite at $q^{\star}$. Hence, from the analysis above, we have the 
following remark.
\begin{remark}
 If $n=m$, the mechanical system \eqref{mecsys} can be stabilized at the desired 
equilibrium with the controllers proposed in Section \ref{sec:int} or via PID-PBC. 
\end{remark}
To understand the effect of the derivative term $-K_{\tt D}\dot{y}$ on a mechanical system, 
consider \eqref{mecsys} in closed-loop with \eqref{upid}-\eqref{pidy}. The resulting 
(cyclo-)passive 
system has a storage function of the form
\begin{equation*}
 H_{\tt pid}(q,p) = T(q,p)+V_{\tt d}(q)+\dfrac{1}{2}\lVert \dot{q}_{\tt a} \rVert^{2}_{K_{\tt D}}  
= 
 V_{\tt d}(q) + \dfrac{1}{2}p^{\top}M_{\tt d}^{-1}(q)p,
\end{equation*} 
where
\begin{equation*}
 V_{\tt d}(q) = V(q)+\dfrac{1}{2}\lVert q_{\tt a}+\kappa 
\rVert^{2}_{K_{\tt I}}; \qquad 
M_{\tt d}(q) = M(q)\left[M(q)+GK_{\tt D}G^{\top} \right]^{-1} M(q)\succ 0. 
\end{equation*} 
Notice that the derivative term modifies the kinetic energy. Notably, considering the identity 
$\dot{q} = M^{-1}(q)p$, we have that
\begin{equation*}
 \dfrac{1}{2}p^{\top}M_{\tt d}^{-1}(q)p = \dfrac{1}{2}p^{\top}M_{\tt d}^{-1}(q)\left[M(q)+GK_{\tt 
D}G^{\top} \right]M_{\tt d}^{-1}(q)p = \dfrac{1}{2}\dot{q}^{\top}\left[M(q)+GK_{\tt 
D}G^{\top} \right]\dot{q}. 
\end{equation*} 
Hence, the derivative term adds mass (inertia) to the system. Fig. \ref{fig:pid} depicts the 
physical interpretation of a fully actuated mechanical system in closed-loop with \eqref{pidy}.

While adding mass (inertia) modifies a mechanical system's transitory behavior, it does not affect 
its stability properties. However, if we can identify multiple storage functions for the same 
system, we can identify multiple passive outputs. This approach can be exploited to modify the 
stability properties of some mechanical systems with the derivative term in PID-PBC. To illustrate 
this, consider a mechanical system of the form \eqref{mecsys}, with
\begin{equation}\label{meccondpid}
\begin{array}{l}
 q\in\rea^{2}; \quad G = \begin{bmatrix}
                         0\\ 1
                        \end{bmatrix}; \quad V(q) = V(q_{\tt u}) \ \text{such that} \ \left( 
\PAR{^{2}V}{q_{\tt u}^{2}} \right)^{\star}< 0; \\[0.15cm]  D(q,p) = \mathbf{0}; \quad M(q) = 
\begin{bmatrix}
                                  m_{\tt uu} & m_{\tt ua} \\ m_{\tt ua} &                    
 
          m_{\tt aa}
                                                                 \end{bmatrix}\in\rea^{2\times 2},
\end{array}
\end{equation} 
where $m_{\tt uu}$ and $m_{\tt aa}$ are positive, and $m_{\tt ua}$ is 
different from 
zero.

Given \eqref{meccondpid}, we need to modify the potential energy corresponding to the unactuated 
coordinate to stabilize the system at $q^{\star}$. Nevertheless, considering $y = \dot{q}_{\tt a}$, 
the controller \eqref{pidy} only modifies the potential energy in the actuated coordinates. To 
overcome this situation, we can identify new storage functions for this system. In particular, from 
\eqref{meccondpid}, $\dot{p}_{\tt a} = u$. Hence,
\begin{equation*}
 H_{\tt a}(p_{\tt a}) = \dfrac{1}{2 m_{\tt aa}} p_{\tt a}^{2}
\end{equation*} 
satisfies
\begin{equation*}
 \dot{H}_{\tt a} = \dfrac{1}{m_{\tt aa}} p_{\tt a}u.
\end{equation*} 
Thus, recalling that $p = M\dot{q}$,
\begin{equation*}
 y_{\tt a} = \dfrac{1}{m_{\tt aa}} p_{\tt a} = \dfrac{m_{\tt ua}}{m_{\tt aa}}\dot{q}_{\tt u} + 
\dot{q}_{\tt a} 
\end{equation*} 
is a passive output. Because $H(q,p)$ and $H_{\tt a}(p_{\tt a})$ are storage functions, any linear 
combination of them yields another storage function. Hence, another storage function is given by
\begin{equation*}
 H_{\tt u}(q,p) = H(q,p) - H_{\tt a}(p_{\tt a}) =  V(q_{\tt u})+ \dfrac{1}{2}p^{\top}M^{-1}p -  
\dfrac{1}{2 m_{\tt aa}} p_{\tt a}^{2}.
\end{equation*} 
Moreover, because $\dot{H} = u\dot{q}_{\tt a}$,
\begin{equation*}
 \dot{H}_{\tt u} = u(\dot{q}_{\tt a}-y_{\tt a})  = -\left( \dfrac{m_{\tt ua}}{m_{\tt 
aa}}\dot{q}_{\tt u} \right)u.
\end{equation*} 
Therefore,
\begin{equation*}
 y_{\tt u} = -\dfrac{m_{\tt ua}}{m_{\tt 
aa}}\dot{q}_{\tt u}.
\end{equation*} 
is a passive output associated with $H_{\tt u}(q,p)$. Exploiting again the fact that any linear 
combination of $H_{\tt u}(q,p)$ and $H_{\tt a}(p_{\tt a})$ yields another storage function, we can 
consider a constant $k_{\tt a}\in\rea$ such that the storage function
\begin{equation*}
 H_{\tt n}(q,p) = k_{\tt a}H_{\tt a}(p_{\tt a})-H_{\tt u}(q,p),
\end{equation*} 
is associated with the passive output
\begin{equation*}
 y_{\tt n} = k_{\tt a}y_{\tt a} - y_{\tt u} = k_{\tt a}\dot{q}_{\tt a} + \dfrac{(k_{\tt 
a}+1)m_{\tt ua}}{m_{\tt a}}\dot{q}_{\tt u}.
\end{equation*} 
Notice that the new passive output $y_{\tt n}$ is integrable, with
\begin{equation*}
 \gamma_{\tt n}(q) =  k_{\tt a}{q}_{\tt a} + \dfrac{(k_{\tt 
a}+1)m_{\tt ua}}{m_{\tt a}}{q}_{\tt u}.
\end{equation*} 
Moreover, $y_{\tt n}$ has relative degree one and some 
lengthy but simple computations show the existence of $k_{\tt a}$ such that the rank condition 
equivalent to \eqref{Kdcond1} is satisfied. Consequently, a full PID passivity-based controller can 
be implemented using this output, in which case, the closed-loop energy function is given by
\begin{equation*}
 H_{\tt pid}(q,p) = H_{\tt n}(q,p)+V^{\star} + \dfrac{K_{\tt I}}{2}\left(\gamma_{\tt 
n}(q)-\gamma_{\tt n}^{\star}  \right)^{2} + \dfrac{K_{\tt D}}{2}y_{\tt n}^{2} = V_{\tt d}(q) + 
T_{\tt d}(p),
\end{equation*} 
with
\begin{equation}\label{VdTd}
 \begin{array}{rlc}
V_{\tt d}(q) &=& -V(q_{\tt u})+V^{\star}+\dfrac{K_{\tt I}}{2}\left(\gamma_{\tt n}(q)-\gamma_{\tt 
n}^{\star}  \right)^{2}; \\[0.2cm] T_{\tt d}(p) &=& \dfrac{1}{2}p^{\top}M^{-1}p +  
\dfrac{k_{\tt a}+1}{2 m_{\tt aa}} p_{\tt a}^{2} +  \dfrac{K_{\tt D}}{2}y_{\tt n}^{2}. \end{array}
\end{equation} 
Note that, from \eqref{eqmec}, $V(q_{\tt u})$ must satisfy
\begin{equation*}
 \left( \displaystyle\PAR{V}{q_{\tt u}} \right)^{\star} = 0.
\end{equation*} 
Furthermore, the sign of $V(q_{\tt u})$ is flipped in $H_{\tt pid}(q,p)$. Hence, $V_{\tt d}(q)$ is 
positive definite at $q_{\star}$ for any positive $K_{\tt I}$. Nevertheless, the sign of the 
open-loop kinetic energy is also flipped. Therefore, enough mass (inertia) must be added through 
the derivative term to guarantee that $T_{\tt d}(p)$ is positive definite at zero, which, given 
\eqref{VdTd}, boils down to choose $k_{\tt a}$ and $K_{\tt D}$ such that\footnote{The computations 
leading to this conclusion are straightforward and are omitted for brevity.}
\begin{equation*}
 \begin{bmatrix}
  K_{\tt D}\left((k_{\tt a}+1)\dfrac{m_{\tt ua}}{m_{\tt aa}}\right)^{2}-m_{\tt uu} & K_{\tt 
D}(k_{\tt 
a}+1)\dfrac{m_{\tt ua}}{m_{\tt aa}}-m_{\tt ua} \\[0.3cm] K_{\tt D}(k_{\tt 
a}+1)\dfrac{m_{\tt ua}}{m_{\tt aa}}-m_{\tt ua} & K_{\tt D}k_{\tt 
a}^{2}+\dfrac{(k_{\tt a}+1)}{m_{\tt aa}}-m_{\tt aa}
 \end{bmatrix}
\succ 0.
\end{equation*} 

A more general formulation of the approach above for mechanical systems in the EL representation is 
exposed in \cite{DONetal,ROMetal} and implemented experimentally in \cite{gandhi}. 

\section{Interconnection and Damping Assignment}
An overview of interconnection and damping assignment (IDA) for general passive systems is provided 
in the chapter Introduction to Passivity-Based Control---see Section \ref{sec:cross}. Here, we 
focus on IDA-PBC for pH systems and its relationship with the PID-PBC approaches described in 
Sections \ref{sec:int} and \ref{sec:pid}.

In IDA-PBC, the desired closed-loop system is a pH system of the form
\begin{equation}\label{clIDA}
 \dot{x} = F_{\tt d}(x)\nabla H_{\tt ida}(x),
\end{equation}
where the matrix $\MAP{F_{\tt d}}{\mathcal{X}}{n\times n}$ satisfies
\begin{equation*}
 F_{\tt d}(x)+F^{\top}_{\tt d}(x)\preceq 0,
\end{equation*} 
while the closed-loop Hamiltonian $H_{\tt ida}(x)$ is positive definite at the desired 
equilibrium $x_{\star}\in\mathcal{E}$. Consequently, $x^{\star}$ is a stable equilibrium for the 
closed-loop system \eqref{clIDA} with Lyapunov function $H_{\tt ida}(x)$. Without loss of 
generality, the matrix $F_{\tt d}(x)$ can be decomposed into $F_{\tt d}(x)=J_{\tt d}(x)-R_{\tt 
d}(x)$, where $J_{\tt d}(x)$ is skew-symmetric and is called the desired interconnection matrix; 
$R_{\tt d}(x)$ is positive definite semi-definite, referred to as the desired dissipation matrix. 
Thus, assigning $F_{\tt d}(x)$ is equivalent to assigning the desired interconnection and damping 
to 
the closed-loop system, hence the method's name \citep{IDAaut}.

Equating the right hand 
of \eqref{phsys} with the right hand of \eqref{clIDA} yields 
\begin{equation}\label{eqIDA}
 F(x)\nabla H(x) + g(x)u = F_{\tt d}(x)\nabla H_{\tt ida}(x).
\end{equation} 
From \eqref{eqIDA}, we can obtain the following expressions:
\begin{eqnarray}\label{meIDA}
 \mathbf{0}&=&g^{\perp}(x)\left\lbrace F_{\tt d}(x)\nabla H_{\tt ida}(x) - F(x)\nabla H(x) 
\right\rbrace;\\
\label{uIDA} u &=& \left[g^{\top}(x)g(x)\right]^{-1}g^{\top}(x)\left\lbrace F_{\tt d}(x)\nabla 
H_{\tt ida}(x)-F(x)\nabla H(x) 
\right\rbrace.
\end{eqnarray} 
The PDE \eqref{meIDA} is often referred to as the matching equation \citep{IDAsurvey}, and finding 
solutions that simultaneously satisfy the restrictions imposed on $F_{\tt d}(x)$ and $H_{\tt 
ida}(x)$ represents the main challenge in the IDA-PBC approach. In this regard, numerous studies 
have focused on proposing constructive methods to solve \eqref{meIDA} or providing closed-form 
solutions for particular classes of systems. For instance, the references 
\cite{nunna,cieza2018ida,pfaffian,borjaUKACC} study the solution to the matching equation for 
general pH systems, while the studies \cite{acosta,viola,SIDA,cieza2019ida,HAR,ARP} focus on 
underactuated mechanical systems in the pH representation.

If a pair $(F_{\tt d}(x),H_{\tt ida}(x))$ solving \eqref{meIDA} is found, then the 
controller that yields the closed-loop dynamics \eqref{clIDA} is given by \eqref{uIDA}. Notably, 
IDA-PBC is more general than the approaches discussed in Sections \ref{sec:int} and \ref{sec:pid} 
in the sense that the control design process---particularly the energy shaping---does not rely on 
finding suitable passive outputs. Notice that, in contrast to PID-PBC, the structure of the 
closed-loop Hamiltonian is not predetermined. Comparisons between IDA-PBC and other PBC techniques 
are discussed in \cite{ORTetaltac08} and \cite{TACBOR21}. To illustrate the relationship 
between IDA-PBC and the methodology studied in Section \ref{sec:int}, assume that $F(x)$ has full 
rank and \eqref{yps} is integrable. Then, set
\begin{equation*}
 F_{\tt d}(x)=F(x), \qquad H_{\tt ida}(x) = H_{\Phi}(x)
\end{equation*} 
in \eqref{clIDA}. Hence, 
\begin{equation*}
\begin{array}{rcl}
  g^{\perp}(x)\left\lbrace F_{\tt d}(x)\nabla H_{\tt ida}(x) - F(x)\nabla H(x) 
\right\rbrace &=& g^{\perp}(x)\left\lbrace F(x)\left[\nabla 
H(x)-F^{-1}(x)g(x)\nabla_{\gamma}\Phi(\gamma(x))\right] - F(x)\nabla H(x) 
\right\rbrace \\ 
&=& g^{\perp}(x)\left\lbrace g(x)\nabla_{\gamma}\Phi(\gamma(x)) 
\right\rbrace\\
&=& \mathbf{0}.
\end{array}
\end{equation*} 
To show the relationship between IDA-PBC and the PID-PBC approach, note that $\mathsf{K}(x)$ 
defined in \eqref{K} satisfies the following
\begin{equation}\label{perpK}
 g^{\perp}(x)\mathsf{K}(x) = g^{\perp}(x) \implies g^{\perp}(x)\mathsf{K}^{-1}(x) = 
g^{\perp}(x)\mathsf{K}(x)\mathsf{K}^{-1}(x) = g^{\perp}(x).
\end{equation} 
Assume that $F(x)$ has full 
rank and \eqref{yps} is integrable. Then, select
\begin{equation*}
 F_{\tt d}(x)=\mathsf{F}(x), \qquad H_{\tt ida}(x) = H_{\tt pid}(x),
\end{equation*} 
with $\mathsf{F}(x)$ and $H_{\tt pid}(x)$ given in \eqref{Fpid} and 
\eqref{Hdpidzero}, respectively. Thus, from \eqref{HI}, \eqref{nHpid}, and \eqref{perpK}, we get 
the following:
\begin{equation*}
\begin{array}{rcl}
  g^{\perp}(x)\left\lbrace F_{\tt d}(x)\nabla H_{\tt ida}(x) - F(x)\nabla H(x) 
\right\rbrace &=& g^{\perp}(x)\left\lbrace \mathsf{F}(x)\nabla H_{\tt pid}(x)- F(x)\nabla H(x) 
\right\rbrace \\ &=&
g^{\perp}(x)\left\lbrace \mathsf{K}^{-1}(x)\left[F(x)-g(x)K_{\tt D}g^{\top}(x) 
\right]\nabla H_{\tt I}(x) - F(x)\nabla H(x)
\right\rbrace\\
&=&
g^{\perp}(x)\left\lbrace F(x)\nabla H_{\tt I}(x) - F(x)\nabla H(x)
\right\rbrace\\ &=&
g^{\perp}(x)\left\lbrace F(x)\nabla H(x) - g(x)K_{\tt I}\left( \gamma(x)+\kappa 
\right)  - F(x)\nabla H(x)
\right\rbrace\\
&=& \mathbf{0}.
\end{array}
\end{equation*} 
The scenarios above illustrate that, under some circumstances, the control methods studied 
in Sections \ref{sec:int} and \ref{sec:pid} represent closed-form solutions to the PDE 
\eqref{meIDA}. Nevertheless, even if the matching equation is solved, the 
mentioned approaches cannot always guarantee that $H_{\tt ida}(x)$ is positive definite at the 
desired equilibrium. Moreover, such closed-form solutions are obtained at the expense of imposing 
additional conditions on the plant, particularly its passive output.

\section{Further Reading}
The well-established control by interconnection PBC technique---not discussed in this 
chapter---is covered in the chapters Introduction to Passivity-Based Control and Port-Hamiltonian 
Nonlinear Systems of this encyclopedia. We also refer the reader to 
\cite{STR,macchelli2006,ORTetaltac08,GEObook,VENVAN,VANJEL,FERMIDDON,IOHD,BORint} for further 
discussion on this topic. 

For a discussion on shifted passivity, we refer the reader to \cite{MONetal,wu2020stabilization}. 
Similarly, the reference \cite{kawano} discusses the so-called 
Krasovskii PBC approach.

PBC results for trajectory tracking are reported in 
\cite{FUJIetaltracking,romero2014globally,CISetal,YAG,Jochem,reyes2022virtual}.

For the discrete-time version of some PBC approaches, we refer the reader to 
\cite{moreschini,mattioni2022discrete,macchelli2023control}
\subsection{Cross-References}\label{sec:cross}
Other chapters of this encyclopedia related to the material exposed above are:
\begin{itemize}
 \item Introduction to Passivity-Based Control
 \item Port-Hamiltonian Nonlinear Systems
\end{itemize}

\section{Summary}
\begin{itemize}
 \item The energy shaping and damping injection processes are key in stabilizing nonlinear systems 
via PBC.
 \item Integrable passive outputs can be exploited for energy shaping.
 \item In PID-PBC, the integral and derivative terms shape the energy of the closed-loop system, 
while the proportional term injects damping.
 \item The relative degree of the passive outputs plays a crucial role in constructing the 
proportional and derivative terms in PID passivity-based controllers.
 \item The linear combination of storage functions can be used to identify new passive outputs, 
enlarging the applicability of PID-PBC.
 \item IDA-PBC does not rely on identifying suitable passive outputs; instead, its applicability 
hinges on solving the matching equation.
 \item In some specific scenarios, the closed-loop dynamics obtained via PID-PBC and other 
PBC approaches are closed-form solutions to the matching equations in IDA-PBC. 
\end{itemize}


\bibliographystyle{plain}
\bibliography{refs_encyclopedia}

@book{VAN,
  title={$L_2$-{G}ain and {P}assivity Techniques in Nonlinear Control},
  author={van der Schaft, A J},
  year={2016},
  publisher={Springer},
  address={Berlin},
  edition={Third}
}

@book{KHA,
  title={Nonlinear Systems},
  author={Khalil, H. K.},
  year={2002},
  publisher = {Prentice-Hall},
  address={New Jersey},
  edition={Third}
}

@article{ORTetaltac08,
  title={Control by interconnection and standard {P}assivity-based control of port-{H}amiltonian 
systems},
  author={Ortega, R. and van der Schaft, A. J. and Casta\~{n}os, F. and Astolfi, A.},
  journal={Automatic Control, IEEE Transactions on},
  volume={53},
  number={11},
  pages={2527--2542},
  year={2008},
  publisher={IEEE}
}

@article{ORTcsm,
	author={Ortega, R. and van der Schaft, A. J. and Mareels, I. and Maschke, B.}, 
	journal={Control Systems Magazine, IEEE}, 
	title={Putting energy back in control}, 
	year={2001}, 
	volume={21}, 
	number={2}, 
	pages={18--33}}

@book{PIDbook,
  title={{PID} Passivity-based Control of Nonlinear Systems with Applications},
  author={Ortega, R. and Romero, J. G. and Borja, P. and Donaire, A.},
  year={2021},
  publisher={John Wiley \& Sons}
}

@article{IDAsurvey,
  title={Interconnection and damping assignment passivity-based control: A survey},
  author={Ortega, R. and Garcia-Canseco, E.},
  journal={European Journal of control},
  volume={10},
  number={5},
  pages={432--450},
  year={2004},
  publisher={Elsevier}
}

@article{IDAaut,
  title={Interconnection and damping assignment passivity-based control of port-controlled 
{H}amiltonian systems},
  author={Ortega, R. and van der Schaft, A. J. and Maschke, B. and Escobar, G.},
  journal={Automatica},
  volume={38},
  number={4},
  pages={585--596},
  year={2002},
  publisher={Elsevier}
}

@article{hill76,
  title={The stability of nonlinear dissipative systems},
  author={Hill, D. and Moylan, P.},
  journal={IEEE transactions on automatic control},
  volume={21},
  number={5},
  pages={708--711},
  year={1976},
  publisher={IEEE}
}

@article{MENetal,
	author={Zhang, M. and Ortega, R. and Jeltsema, D. and Su, H.},
	title={Further deleterious effects of the dissipation obstacle in control by interconnetction of port-{H}amiltonian systems},
	journal={Automatica},
	volume={25},
	number={6},
	pages={877--888},
	year={2015},
	}

@article{MENBOORT,
  title={{PID} passivity-based control of port-{H}amiltonian systems},
  author={Zhang, M. and Borja, P. and Ortega, R. and Liu, Z. and Su, H.},
  journal={IEEE Transactions on Automatic Control},
  volume={63},
  number={4},
  pages={1032--1044},
  year={2017},
  publisher={IEEE}
	}

@article{VANJEL,
	author={van der Schaft, A. J. and Jeltsema, D.},
	journal={Foundations and Trends in Systems and Control},
	title={Port-{H}amiltonian systems theory: an introductory overview},
	volume={1},
	number={2-3},
	pages={173-378},
	year={2014},
	}

@book{GEObook,
  title={{M}odeling and control of complex physical systems: the port-{H}amiltonian approach},
  author={Duindam, V. and Macchelli, A. and Stramigioli, S. and Bruyninckx, H.},
  year={2009},
  publisher={Springer Science \& Business Media}
}

@article{kawano,
  title={Krasovskii and shifted passivity-based control},
  author={Kawano, Y. and Kosaraju, K. C. and Scherpen, J. M. A.},
  journal={IEEE Transactions on Automatic Control},
  volume={66},
  number={10},
  pages={4926--4932},
  year={2020},
  publisher={IEEE}
}

@article{ROMetal,
  title={Global stabilisation of underactuated mechanical systems via {PID} passivity-based control},
  author={Romero, J. G. and Donaire, A. and Ortega, R. and Borja, P.},
  journal={Automatica},
  volume={96},
  pages={178--185},
  year={2018},
  publisher={Elsevier}
}

@article{gandhi,
  title={Energy shaping control of an inverted flexible pendulum fixed to a cart},
  author={Gandhi, P. S. and Borja, P. and Ortega, R.},
  journal={Control Engineering Practice},
  volume={56},
  pages={27--36},
  year={2016},
  publisher={Elsevier}
}

@ARTICLE{TACBOR21,
  author={P. {Borja} and R. {Ortega} and J. M. A. {Scherpen}},
  journal={IEEE Transactions on Automatic Control},
  title={New Results on Stabilization of port-{H}amiltonian Systems via {PID} Passivity-based Control},
  year={2021},
  volume={66},
  number={2},
  pages={625-636}}

@article{DONetal,
	author={Donaire, A. and Mehra, R. and Ortega, R. and Satpute, S. and Romero, J. G. and Kazi, F. and Singh, N. M.},
	journal={Automatic Control, IEEE Transactions on},
	title={Shaping the energy of mechanical systems without solving partial differential equations},
	volume = {6},
	number = {8},
	pages = {1051--1056},
	year = {2016}}

@book{ORTbook,
	author    = "Ortega, R. and Lor\'{\i}a, A. and Nicklasson, P. J. and Sira-Ram{\'\i}rez, 
H.",
	title     = "{P}assivity-{B}ased {C}ontrol of {E}uler-{L}agrange {S}ystems: {M}echanical, 
{E}lectrical and {E}lectromechanical {A}pplications",
	year      = "1998",
	publisher = "Communications and Control Engineering. Springer Verlag, London"}

@article{MONetal,
  title={Conditions on shifted passivity of port-{H}amiltonian systems},
  author={Monshizadeh, N. and Monshizadeh, P. and Ortega, R. and van der Schaft, A. J.},
  journal={Systems \& Control Letters},
  volume={123},
  pages={55--61},
  year={2019},
  publisher={Elsevier}
}

@article{CISetal, 
	author={Cisneros, R. and Pirro, M. and Bergna, G. and Ortega, R. and Ippoliti, G. and Molinas, 
M.}, 
	journal={Control Engineering Practice}, 
	title={Global tracking passivity-based {PI} control of bilinear systems and its application to 
the boost and modular multilevel converters}, 
	volume={43},
	pages={109--119},
	year={2015}
	}

@INPROCEEDINGS{FERMIDDON, 
	author={Ferguson, J. and Middleton, R. H. and Donaire, A.}, 
	booktitle={54th IEEE Conference on Decision and Control}, 
	title={Disturbance rejection via control by interconnection of port-{H}amiltonian systems}, 
	year={2015}, 
	month={Dec},  
	pages={507-512}}

@inproceedings{macchelli2006,
  title={Control of port {H}amiltonian systems by interconnection and energy shaping via generation 
of {C}asimir functions. An overview},
  author={Macchelli, Alessandro and Pasumarty, R and van der Schaft, A},
  booktitle={Proc. 5th Int. Symp. Mathematical Modeling (MATHMOD)},
  pages={5--1},
  year={2006},
  organization={Vienna, Austria}
}

@article{VENVAN,
  title={Energy shaping of port-{H}amiltonian systems by using alternate passive input-output 
pairs},
  author={Venkatraman, A. and van der Schaft, A. J.},
  journal={European Journal of Control},
  volume={16},
  number={6},
  pages={665--677},
  year={2010},
  publisher={Elsevier}
}

@inproceedings{IOHD,
  title={Interconnections of input-output {H}amiltonian systems with dissipation},
  author={van der Schaft, A. J.},
  booktitle={2016 IEEE 55th Conference on Decision and Control (CDC)},
  pages={4686--4691},
  year={2016},
  organization={IEEE}
}

@article{STR,
  title={Passive output feedback and port interconnection},
  author={Stramigioli, S. and Maschke, B. and van der Schaft, A. J.},
  journal={IFAC Proceedings Volumes},
  volume={31},
  number={17},
  pages={591--596},
  year={1998},
  publisher={Elsevier}
}

@article{BORint,
  title={Interconnection Schemes in Modeling and Control},
  author={Borja, P. and Ferguson, J. and van der Schaft, A. J.},
  journal={IEEE Control Systems Letters},
  year={2023},
  publisher={IEEE}
}

@article{acosta,
  title={Interconnection and damping assignment passivity-based control of mechanical systems with 
underactuation degree one},
  author={Acosta, J. A. and Ortega, R. and Astolfi, A. and Mahindrakar, A. D.},
  journal={IEEE Transactions on Automatic Control},
  volume={50},
  number={12},
  pages={1936--1955},
  year={2005},
  publisher={IEEE}
}

@article{nunna,
  title={Constructive interconnection and damping assignment for port-controlled {H}amiltonian 
systems},
  author={Nunna, K. and Sassano, M. and Astolfi, A.},
  journal={IEEE Transactions on Automatic Control},
  volume={60},
  number={9},
  pages={2350--2361},
  year={2015},
  publisher={IEEE}
}

@article{cieza2018ida,
  title={{IDA}-{PBC} for polynomial systems: An {SOS}-based approach},
  author={Cieza, O. B. and Reger, J.},
  journal={IFAC-PapersOnLine},
  volume={51},
  number={13},
  pages={366--371},
  year={2018},
  publisher={Elsevier}
}

@inproceedings{cieza2019ida,
  title={{IDA}-{PBC} for underactuated mechanical systems in implicit port-{H}amiltonian 
representation},
  author={Cieza, O. B. and Reger, J.},
  booktitle={2019 18th European Control Conference (ECC)},
  pages={614--619},
  year={2019},
  organization={IEEE}
}

@article{pfaffian,
	title = {Solution of matching equations of {IDA}-{PBC} by {Pfaffian} differential equations},
	volume = {95},
	number = {12},
	journal = {International Journal of Control},
	author = {Harandi, M. R. J. and Taghirad, H. D.},
	year = {2022},
	pages = {3368--3378}
}

@article{ARP,
	title = {A Constructive Methodology for the {IDA}-{PBC} of Underactuated 2-{DoF} Mechanical 
Systems with Explicit Solution of {PDEs}},
	volume = {20},
	language = {en},
	number = {1},
	journal = {International Journal of Control, Automation and Systems},
	author = {Arpenti, P. and Ruggiero, F. and Lippiello, V.},
	year = {2022},
	pages = {283--297}
}

@article{viola,
  title={Total energy shaping control of mechanical systems: simplifying the matching equations via 
coordinate changes},
  author={Viola, G. and Ortega, R. and Banavar, R. and Acosta, J. A. and Astolfi, A.},
  journal={IEEE Transactions on Automatic Control},
  volume={52},
  number={6},
  pages={1093--1099},
  year={2007},
  publisher={IEEE}
}

@article{SIDA,
  title={Simultaneous interconnection and damping assignment passivity-based control of mechanical 
systems using dissipative forces},
  author={Donaire, A. and Ortega, R. and Romero, J. G.},
  journal={Systems \& Control Letters},
  volume={94},
  pages={118--126},
  year={2016},
  publisher={Elsevier}
}

@article{HAR,
	title = {On the matching equations of kinetic energy shaping in {IDA}-{PBC}},
	volume = {358},
	number = {16},
	urldate = {2023-11-14},
	journal = {Journal of the Franklin Institute},
	author = {Harandi, M. R. J. and Taghirad, H. D.},
	month = oct,
	year = {2021},
	pages = {8639--8655}
}

@article{YAG,
  title={Trajectory tracking for a class of contractive port {H}amiltonian systems},
  author={Yaghmaei, A. and Yazdanpanah, M. J.},
  journal={Automatica},
  volume={83},
  pages={331--336},
  year={2017},
  publisher={Elsevier}
}

@article{ortega1989adaptive,
  title={Adaptive motion control of rigid robots: A tutorial},
  author={Ortega, R. and Spong, M. W.},
  journal={Automatica},
  volume={25},
  number={6},
  pages={877--888},
  year={1989},
  publisher={Elsevier}
}

@article{WIL,
  title={Dissipative dynamical systems part {I}: General theory},
  author={Willems, J. C.},
  journal={Archive for rational mechanics and analysis},
  volume={45},
  number={5},
  pages={321--351},
  year={1972},
  publisher={Springer}
}

@article{moreschini,
  title={Stabilization of discrete port-{H}amiltonian dynamics via interconnection and damping 
assignment},
  author={Moreschini, A. and Mattioni, M. and Monaco, S. and Normand-Cyrot, 
D.},
  journal={IEEE Control Systems Letters},
  volume={5},
  number={1},
  pages={103--108},
  year={2020},
  publisher={IEEE}
}

@article{FUJIetaltracking,
  title={Trajectory tracking control of port-controlled {H}amiltonian systems via generalized canonical transformations},
  author={Fujimoto, K. and Sakurama, K. and Sugie, T.},
  journal={Automatica},
  volume={39},
  number={12},
  pages={2059--2069},
  year={2003},
  publisher={Elsevier}
}

@article{BYR91,
  title={Passivity, feedback equivalence, and the global stabilization of minimum phase nonlinear systems},
  author={Byrnes, C. I. and Isidori, A. and Willems, J. C.},
  journal={IEEE Transactions on automatic control},
  volume={36},
  number={11},
  pages={1228--1240},
  year={1991}
}

@article{chan2023dead,
  title={Dead-zone compensation via passivity-based control for a class of mechanical systems},
  author={Chan-Zheng, C. and Borja, P. and Scherpen, J.M.A.},
  journal={IFAC-PapersOnLine},
  volume={56},
  number={1},
  pages={319--324},
  year={2023},
  publisher={Elsevier}
}

@article{borja2023stabilization,
  title={Stabilization of physical systems via saturated controllers with partial state 
measurements},
  author={Borja, P. and Chan-Zheng, C. and Scherpen, J.M.A.},
  journal={IEEE Transactions on Control Systems Technology},
  volume={31},
  number={6},
  pages={2405--2419},
  year={2023},
  publisher={IEEE}
}

@inproceedings{BORDASAN,
  title={Energy-based shape regulation of soft robots with unactuated dynamics dominated by 
elasticity},
  author={Borja, P. and Dabiri, A. and Della Santina, C.},
  booktitle={2022 IEEE 5th International Conference on Soft Robotics (RoboSoft)},
  pages={396--402},
  year={2022},
  organization={IEEE}
}

@inproceedings{borjaUKACC,
  title={Interconnection and Damping Assignment Passivity-Based Control Without Partial Differential 
Equations},
  author={Borja, P.},
  booktitle={2024 UKACC 14th International Conference on Control (CONTROL)},
  pages={131--136},
  year={2024},
  organization={IEEE}
}

@article{takegaki,
  author={Takegaki, M. and Arimoto, S.},
     title = {A New Feedback Method for Dynamic Control of Manipulators},
    journal = {Journal of Dynamic Systems, Measurement, and Control},
    volume = {103},
    number = {2},
    pages = {119-125},
    year = {1981}
}

@book{bai,
  title={Cooperative control design: a systematic, passivity-based approach},
  author={Bai, H. and Arcak, M. and Wen, J.},
  year={2011},
  publisher={Springer Science \& Business Media}
}

@book{secchi,
  title={Control of interactive robotic interfaces: A port-{H}amiltonian approach},
  author={Secchi, C. and Stramigioli, S. and Fantuzzi, C.},
  volume={29},
  year={2007},
  publisher={Springer Science \& Business Media}
}

@article{brogliato,
  title={Dissipative systems analysis and control},
  author={Brogliato, B. and Lozano, R. and Maschke, B. and Egeland, O.},
  journal={Theory and Applications},
  volume={2},
  pages={2--5},
  year={2007},
  publisher={Springer}
}

@article{jeltsema2009multidomain,
  title={Multidomain modeling of nonlinear networks and systems},
  author={Jeltsema, D. and Scherpen, J. M. A.},
  journal={IEEE Control Systems Magazine},
  volume={29},
  number={4},
  pages={28--59},
  year={2009},
  publisher={IEEE}
}

@article{wu2020stabilization,
  title={Stabilization of port-{H}amiltonian systems based on shifted passivity via feedback},
  author={Wu, C. and van der Schaft, A. J. and Chen, J.},
  journal={IEEE Transactions on Automatic Control},
  volume={66},
  number={5},
  pages={2219--2226},
  year={2020},
  publisher={IEEE}
}

@article{romero2014globally,
  title={A globally exponentially stable tracking controller for mechanical systems using position 
feedback},
  author={Romero, J. G. and Ortega, R. and Sarras, I.},
  journal={IEEE Transactions on Automatic Control},
  volume={60},
  number={3},
  pages={818--823},
  year={2014},
  publisher={IEEE}
}

@INPROCEEDINGS{Jochem,
  author={Borja, P. and van der Veen, J. and Scherpen, J. M. A.},
  booktitle={2021 60th IEEE Conference on Decision and Control (CDC)}, 
  title={Trajectory Tracking for Robotic Arms with Input Saturation and Only Position 
Measurements}, 
  year={2021},
  volume={},
  number={},
  pages={2434-2439}
  }

@article{reyes2022virtual,
  title={Virtual contractivity-based control of fully-actuated mechanical systems in the 
port-{H}amiltonian framework},
  author={Reyes-B{\'a}ez, R. and van der Schaft, A. J. and Jayawardhana, B.},
  journal={Automatica},
  volume={141},
  pages={110275},
  year={2022},
  publisher={Elsevier}
}

@article{mattioni2022discrete,
  title={Discrete-time energy-balance passivity-based control},
  author={Mattioni, M. and Moreschini, A. and Monaco, S. and Normand-Cyrot, 
D.},
  journal={Automatica},
  volume={146},
  pages={110662},
  year={2022},
  publisher={Elsevier}
}

@article{macchelli2023control,
  title={Control design for a class of discrete-time port-{H}amiltonian systems},
  author={Macchelli, A.},
  journal={IEEE Transactions on Automatic Control},
  year={2023},
  publisher={IEEE}
}

\end{document}